\documentclass[runningheads]{llncs}

\usepackage[T1]{fontenc}

\def\doi#1{\href{https://doi.org/\detokenize{#1}}{\url{https://doi.org/\detokenize{#1}}}}

\usepackage{graphicx}
\usepackage{algorithm2e}
\usepackage{amsfonts}

\usepackage{glossaries}

\usepackage{listings}
\usepackage{xcolor}

\definecolor{codegreen}{rgb}{0,0.6,0}
\definecolor{codegray}{rgb}{0.5,0.5,0.5}
\definecolor{codepurple}{rgb}{0.58,0,0.82}
\definecolor{backcolour}{rgb}{0.95,0.95,0.92}

\lstdefinestyle{mystyle}{
    backgroundcolor=\color{backcolour},   
    commentstyle=\color{codegreen},
    keywordstyle=\color{magenta},
    numberstyle=\tiny\color{codegray},
    stringstyle=\color{codepurple},
    basicstyle=\ttfamily\footnotesize,
    breakatwhitespace=false,         
    breaklines=true,                 
    captionpos=b,                    
    keepspaces=true,                 
    numbers=left,                    
    numbersep=5pt,                  
    showspaces=false,                
    showstringspaces=false,
    showtabs=false,                  
    tabsize=2
}

\usepackage{todonotes}

\newcommand{\colvec}[2]{\begin{bmatrix}#1\\#2\end{bmatrix}}

\usepackage{cleveref}
\usepackage{enumitem}
\usepackage{amsmath}

\begin{document}

\title{A Validation Procedure for the Estimation of Reachable Regions in Football}

\author{Moritz Renkin\inst{1}
\and Jonas Bischofberger\inst{2}
\and Erich Schikuta\inst{1}
\and Arnold Baca\inst{2}}

\authorrunning{M. Renkin et al.}

\institute{Research Group Workflow Systems and Technology, Faculty of Computer Science, University of Vienna, Austria\\\email{a11807211@unet.univie.ac.at} \and
Department of Biomechanics, Kinesiology and Computer Science in Sport, Institute of Sports Science, University of Vienna, Austria\\\email{jonas.bischofberger@univie.ac.at}}

\maketitle

\begin{abstract}
Modelling the trajectorial motion of humans along the ground is a foundational task in the quantitative analysis of sports like association football. Most existing models of football player motion have not been validated yet with respect to actual data. One of the reasons for this lack is that performing such a validation is not straightforward, because models of player motion are usually phrased in a way that emphasises possibly reachable positions rather than expected positions. Since positional data of football players typically contains outliers, this data may misrepresent the range of actually reachable positions.

This paper proposes a validation routine for trajectorial motion models that measures and optimises the ability of a motion model to accurately predict all possibly reachable positions by favoring the smallest predicted area of reachable positions that encompasses all observed reached positions up to a manually defined threshold. We demonstrate validation and optimisation on four different motion models, assuming (a) motion with constant speed, (b) motion with constant acceleration, (c) motion with constant acceleration with a speed limit, and (d) motion along two segments with constant speed. Our results show that assuming motion with constant speed or constant acceleration without a limit on the achievable speed is particularly inappropriate for an accurate distinction between reachable and unreachable locations. Motion along two segments of constant speed provides by far the highest accuracy among the tested models and serves as an efficient and accurate approximation of real-world player motion.

\end{abstract}

\section{Introduction}

Recently, professional association football has seen a surge in the availability of positional data of the players and the ball, typically collected by GPS, radar or camera systems~\cite{pino2021comparison}. The growing availability of such data has opened up an exciting new avenue for performance analysis~\cite{memmert2018data}. Positional data enables the development of sophisticated performance indicators that accurately measure the technical, tactical and athletic performance of teams and players beyond the possibilities of qualitative observation and simple event-based statistics like ball possession and passing accuracy. High-quality measures of performance are invaluable for effective performance analysis, training, opposition scouting, and player recruitment.

The modelling of human motion is a foundational component of many performance metrics based on positional data. For example, algorithms that compute space control~\cite{rico2020identification} or simulate passes~\cite{spearman2017physics} implicitly or explicitly make assumptions about human kinematics. These kinematic assumptions have never been verified so far, which calls the validity of these assumptions and the resulting models into question.

In sports with many degrees of freedom like football, predicting player motion is generally a very hard task because player positioning and motion are the result of an intractable, individual decision-making process. Luckily, many applications do not require predictions of the actual position and kinematic state of players but merely of their possibly reachable positions. This requirement essentially shifts the purpose of a motion model from predicting expected positions towards estimating the most remote reachable positions. Estimating such extreme values from real-world data can be difficult, because extreme values are typically rare and particularly likely to include a component of measurement error. Since the distribution of measurement error within player trajectories can vary significantly between data sets due to the use of different measurement systems and post-processing methods, a validation procedure of player motion models needs to flexibly account for various, possibly unknown distributions of measurement error.

The contributions of this paper are twofold: First, we formally propose a validation routine for the quality of player motion models which measures their ability to predict all reachable positions depending on the player's current position and kinematic state. The procedure favors those models that predict the smallest reachable areas which still contain a certain, manually defined proportion of actually observed positions. Second, we use this routine to evaluate and optimise the parameters of four models of motion, assuming (a) motion with constant speed, (b) motion with constant acceleration, (c) motion with constant acceleration with a speed limit, and (d) motion along two segments of constant speed. This evaluation sheds light on the predictive quality of these models and suggests sensible parameter values for them.

The rest of this paper is structured as follows: Section~\ref{state_of_the_art} provides some background on motion models in football and their validation. Section~\ref{validation_procedure} formally presents our validation routine. Section~\ref{experiment_results} describes our exemplary model validation and optimisation based on a real data set and discusses its results. Section~\ref{conclusion} summarises the contributions of this paper and points out possible directions of further research.

\section{Motion models in football: state of the art} \label{state_of_the_art}

Assumptions about the trajectorial motion of players are inherent to many performance indicators within the analysis of sports games. One example is the commonly used concept of space control which assigns control or influence over different areas on the pitch to players. It is used, for example, as a context variable to rate football actions~\cite{fernandez2018wide,rein2017pass} and for time series analyses~\cite{fonseca2012spatial}. Controlled space is often defined as the area that a player is able to reach before any other player, given a specific model of motion for each player. Commonly used for this purpose are motion models assuming constant and equal speed, which results in a Voronoi partition of the pitch~\cite{efthimiou2021voronoi}, or accelerated movement, possibly including a friction term~\cite{fujimura2005} or velocity-dependent acceleration~\cite{taki2000visualization} to limit the achievable speed. Spearman et al.~\cite{spearman2017physics} assume accelerated player motion with a limit on both acceleration and velocity in the context of modeling ground passes.

Motion models have also been estimated directly from positional data by fitting a probability distribution over appropriately normalized future positions~\cite{brefeld2019probabilistic,caetano2021football}. However, such empirical models can be computationally expensive, prone to outliers and their current versions lend themselves less naturally to extreme value estimation than theoretically derived models.
Attempts to validate trajectorial player motion models are rare. Notably, Caetano et al.~\cite{caetano2021football} performed a validation of their space control model, and thus indirectly also the underlying motion model, by checking how many future positions of players fall within their associated controlled area for a number of time horizons.

\section{Player motion model and validation procedure} \label{validation_procedure}
\subsection{Objectives \& Requirements} \label{objectives}
The essential requirement for a player motion model with regards to our validation routine is that it defines a non-zero, finite area that corresponds to the set of positions that the player is able to reach according to the model.

The validation procedure can be represented as a function rating a suitable player motion model on how well it fits some real positional data. Usually, positional data in soccer consists of individual frames, with the position of each player defined for each frame. The frames are normally separated by a constant amount of time (seconds per frame).
In order to abstract our validation procedure from the underlying positional data, we introduce the concept of a \textit{trail}. A trail represents a slice of a player's trajectory over some duration $\Delta t$. Formally, a trail is defined as the quadruple:
\begin{equation} \label{eq_trail_def}
    (\vec{x}_0, \vec{v}_0, \vec{x}_t, \Delta t)
\end{equation}

\begin{itemize}
    \item $\vec{x}_0$: (2D) position of a given player at some arbitrary time $t_0$
    \item $\vec{v}_0$: (2D) velocity of the player at time $t_0$ 
    \item $\vec{x}_t$:  (2D) position of the player at time $t = t_0 + \Delta t$
    \item $\Delta t$: time horizon (predefined)
\end{itemize}
\begin{figure} \label{fig:trail}
    \centering
    \includegraphics[width=260pt]{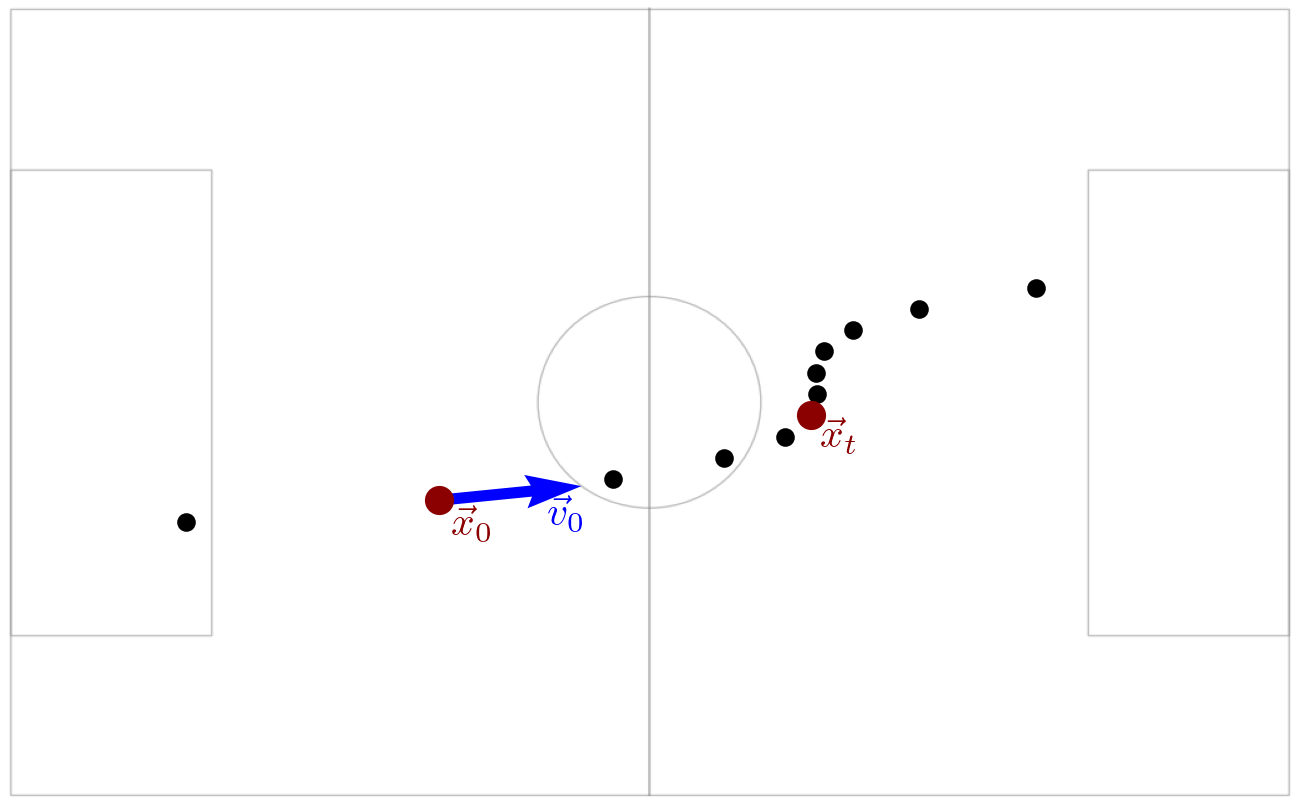}
    \caption{A trail visualized. The black dots represent the positions of a player for a number of consecutive frames. In this case, the time horizon $\Delta t$ is four times the duration of a frame. The player's current velocity $\vec{v}_0$ can be calculated using numerical differentiation if it is not present in the data.}
    \label{fig:trail}
\end{figure}
Figure \ref{fig:trail} visualizes how a trail can be extracted from the positional data of a player.

The validation function takes a motion model $m$ and a set of trails $T$ as parameters and returns a numerical validation score. Formally, the function has the signature: $(m, T) \rightarrow \mathbb{R}$.

Since every reached position is trivially contained in a large enough area, the validation function should take not only correctness but also precision of the model into account. The correctness of a motion model measures its ability to make true predictions, i.e. to predict reachable areas that contain the true target position $\vec{x}_t$. Precision refers to how well narrowed-down the predicted areas of a model are. There is a trade-off relationship between correctness and precision.

One non-functional requirement worth mentioning is performance. Since a player motion model usually has to be invoked and evaluated separately for every trail, the computational complexity is generally $\theta(n)$ with $n$ being the number of trails, unless some algorithmic optimisation or approximation is applied. Given that a high number of trails is desirable in order to get a representative sample of player displacements including a sufficient number of extrema, the validation of a trail should be as efficient as possible.

\subsection{Conceptual approach} \label{conceptual_approach}

\subsubsection{Measuring Correctness.} For the conceptual understanding of the correctness measure, it helps to consider only a single trail \eqref{eq_trail_def}  of a player.
Using $\vec{x}_0$, $\vec{v}_0$ and $\Delta t$, a motion model makes a prediction for the reachable area. The reachable area predicted by a motion model $m$ is defined by the set of reachable positions $R_m$. If $\vec{x}_t$ is contained in $R_m$, the model $m$ has made a correct prediction. Following this logic, a motion model achieves the highest possible correctness if and only if for every trail, the model predicts a reachable area in which $\vec{x}_t$ is contained. As in practice, there may appear considerable outliers due to measurement errors, we decided against weighting incorrect predictions according to their distance to the predicted reachable area. 
The ratio between the number of correct predictions $n_{correct}$ and the number of total predictions $n_{total}$ of a model $m$ for a sample of trails $T$ will be called $hit\_ratio$.
\begin{equation} \label{eq_hit_ratio}
    hit\_ratio(m, T) = \cfrac{n_{correct}}{n_{total}} = \cfrac{n_{correct}}{length(T)}
\end{equation}
The total number of predictions is always equal to the length of the sample of trails as the motion model makes exactly one prediction for each trail.
We can use the $hit\_ratio$ of a model as an indicator for its correctness. A high $hit\_ratio$ corresponds to a high correctness and vice-versa.

\subsubsection{Measuring Precision.}
In the context of this paper, the precision of a motion model represents how much it narrows down the reachable area of a player. Smaller reachable areas imply a higher precision of the model and are generally preferable, given an equal $hit\_ratio$. 

To determine the precision of a model across multiple evaluated trails, we use the inverse of the mean surface area of all correctly predicted reachable areas. Incorrect predictions, where the target position $\vec{x_t}$ is not contained in the predicted reachable area are excluded from this average, since the precision of a model would otherwise increase inappropriately for very narrow, incorrect predictions. The precision of model $m$ across a sample of trails $T$ is given by:

\begin{equation} \label{eq_precision}
    precision(m, T) = \cfrac{1}{\cfrac{1}{n_{correct}} \cdot \sum areas_{correct}} = \cfrac{n_{correct}}{\sum areas_{correct}}
\end{equation}
where $\sum areas_{correct}$ is the sum of all correctly predicted reachable areas. %by $m$ for the Trail sample $T$.

\subsubsection{Defining an overall Validation Score.} Since we aim for a single numerical value as a score for player motion models, correctness and precision have to be balanced in some way. Due to the fact that some measurement-related extreme outliers can usually be expected in positional data from football games, a model with a $hit\_ratio$ of 100\% might not necessarily be desirable. Therefore, we introduce a minimum level of correctness $hit\_ratio_{min}$, which represents a minimal required ratio between correct and total predictions of a model. We propose that if a motion model $m$ satisfies the condition 
\[hit\_ratio(m,T) \geq hit\_ratio_{min}\]
for a trail sample $T$, the exact $hit\_ratio(m,T)$ should be indifferent for the overall validation score of $m$. This way, extreme outliers in the positional data caused by measurement-related errors have no influence on the validation score, as long as $hit\_ratio_{min}$ is chosen adequately.%\footnote{This assumes that those extreme outliers cause an incorrect prediction of $m$.}

Consequently, for a motion model $m$ that exceeds $hit\_ratio_{min}$, the validation score is only determined by the precision of the model \eqref{eq_precision}. We define the $score$ of a motion model $m$ with the sample of trails $T$ as:

\begin{equation} \label{eq_validation_score}
    score(m,T) = 
    \begin{cases}
        0 & \text{if $hit\_ratio(m,T) < hit\_ratio_{min}$}\\
        precision(m,T) & \text{else}
    \end{cases}
\end{equation}

The $score$ measures how well a motion models fits a sample of positional data. $hit\_ratio_{min}$ can be considered a free parameter of this validation procedure. Although the primary rationale behind introducing this parameter is to prevent outliers from influencing the validation score, it can also be tuned to some extent for adjusting the behaviour of the validation procedure. Using a relatively high $hit\_ratio_{min}$ favors models with a high correctness, whereas a lower one favors models with a high precision. It is important to note that due to this trade-off relationship, the "right" balance between correctness and precision depends on the specific use-cases of the model. Thus, the exact value for $hit\_ratio_{min}$ has to be chosen based on both the expected quality of the positional data in terms of outliers and the desired properties of the model with regard to correctness vs. precision.

In any case, $hit\_ratio_{min}$ has to be be chosen such that
\begin{equation} \label{eq_condition_hit_ratio_min}
     hit\_ratio_{min} \leq 1 - \cfrac{n_{outlier}}{length(T)}
\end{equation}
where $n_{outlier}$ is the (expected) number of outliers in the sample T. If \eqref{eq_condition_hit_ratio_min} were violated, the validation score would be influenced by the outliers in the sample, which invalidates the primary purpose of using $hit\_ratio_{min}$ as a threshold for the validation score. The identification of outliers or estimation of their prevalence within a sample is naturally non-trivial, since the appropriate statistical method for achieving an acceptable estimate depends on the specific properties of the sample. However, this paper will not go into further detail here.

\subsection{Implementation}
For the actual computation of the validation score of a given motion model with reasonable efficiency, we need to overcome a few challenges. This section outlines how these challenges can be managed when implementing the validation procedure.
\subsubsection{Implementing a motion model interface.}
First of all, a player motion model $m$ predicts a set of reachable positions $R_m$ based on the input parameters. This set $R_m$, however, is hardly useful for implementing our validation procedure because it is an infinite set. In practice, $R_m$ typically corresponds to a bounded, simply-connected shape, so $R_m$ is also defined through its boundary $B_m$. We can use a simple polygon to approximate $B_m$ and thus $R_m$. With an increasing number of polygon vertices, this approximation can become arbitrarily accurate.

Using a polygon to represent the reachable area predicted by a model does not only allow a straightforward computation of its validation score but is also practical for defining a common interface for motion models. Therefore, we implement the abstract concept of a motion model as an interface with the parameters $(\vec{x}_0, \vec{v}_0, \Delta t)$ and an array of vertices representing a polygon as return type.

\subsubsection{Implementing the validation function.} Translating the conceptual approach for calculating the validation score is fairly straightforward. Both the $hit\_ratio$ \eqref{eq_hit_ratio} and the $precision$ \eqref{eq_precision} have to be calculated.

It has to be determined for each trail in $T$ whether $\vec{x}_t$ is inside the polygon returned by the motion model for the parameters $(\vec{x}_0, \vec{v}_0, \Delta t)$ of the current trail. For all trails where this is the case, it is also necessary to calculate the surface area of that polygon in order to later compute the precision \eqref{eq_precision}.

However, looping over the entire sample of trails is not always necessary. Following the definition of the validation score \eqref{eq_validation_score}, the score is always equal to 0 if the condition $hit\_ratio(m,T) < hit\_ratio_{min}$ is true. We can reformulate the hit ratio \eqref{eq_hit_ratio} as:
\[
    hit\_ratio(m, T) = \cfrac{n_{correct}}{length(T)} = \cfrac{length(T)-n_{incorrect}}{length(T)}
\]
with $n_{incorrect}$ being the number of incorrect predictions made by $m$ for $T$. We can put this reformulated definition of $hit\_ratio(m,T)$ into our condition from the definition of the validation score \eqref{eq_validation_score}:
\[
    \cfrac{length(T)-n_{incorrect}}{length(T)} < hit\_ratio_{min}
\]
Solved for $n_{incorrect}$:
\begin{equation} \label{eq_n_incorrect}
    n_{incorrect} > (1-hit\_ratio_{min}) \cdot length(T) 
\end{equation}
Therefore, $score(m, T)$ is equal to 0 if and only if this condition \eqref{eq_n_incorrect} is satisfied. 
As $length(T)$ and $hit\_ratio_{min}$ are known beforehand, the execution of the validation procedure can be stopped prematurely if the number of incorrect predictions exceeds the threshold $(1-hit\_ratio_{min}) \cdot length(T)$. This small optimisation to the original procedure drastically accelerates the validation of motion models with a low correctness, i.e. a low $hit\_ratio$. 

Algorithm \ref{alg:validation} describes the implementation for the computation of the validation score including the mentioned optimisation.
\noindent \begin{algorithm}
  \SetKwInOut{Input}{inputs}
  \SetKwInOut{Output}{output}
  \SetKwProg{validate}{validate}{}{}

%\validate{$(m,T,hit\_ratio_{min})$}{
    \Input{motion model $m$, sample of trails $T$, free parameter $hit\_ratio_{min}$}
    \Output{The validation score as defined in \eqref{eq_validation_score}}
    \BlankLine
    $n_{incorrect} \gets 0$\;
    Let $A$ be an empty array\;
    Let $length(T)$ be the number of trails in $T$\;
    \ForEach{$trail \in T$}{
      \If{$n_{incorrect} > (1-hit\_ratio_{min})length(T)$}{
            \KwRet{0}
        }
        Let $(\vec{x}_0, \vec{v}_0, \vec{x}_t, \Delta t)$ represent the current $trail$\;
        Let $P$ be the reachable area polygon predicted by $m$ based on $(\vec{x}_0,\vec{v}_0,\Delta t)$\;
        \eIf{$\vec{x}_t$ is contained in $P$}{
            Let $a$ be the surface area of $P$\;
            Append $a$ to $A$\;
        }{
            $n_{incorrect} \gets n_{incorrect} + 1$\;
        }
    }
    $n_{correct} \gets length(T) - n_{incorrect}$\;
    \KwRet{$n_{correct}/sum(A)$}\;
    \BlankLine
%}
\caption{Validation routine, optimized}\label{alg:validation}
\end{algorithm}
\section{Experiment \& Evaluation of results} \label{experiment_results}

To illustrate validation and parameter optimisation using the procedure defined in section~\ref{validation_procedure}, we evaluate four different models of motion, assuming (a) motion with constant speed, (b) motion with constant acceleration, (c) motion with constant acceleration until a speed limit is reached, and (d) motion along two segments with constant speed.

\subsection{Data set}

For the evaluation, we use the public sample data set provided by Metrica Sports which consists of three anonymised games of football~\cite{metrica}. The positional data has been collected using a video-based system and is provided at a frequency of 25 Hz. Since the data contains no velocity, we compute a player's velocity $\vec{v}_i$ for each frame $i$ as $\vec{v}_i = \frac{\vec{x}_{i+1} - \vec{x}_{i-1}}{2 \cdot 0.04s}$.

For performing our validation routine we convert the positional data into a list of trails. As outlined in \eqref{eq_trail_def}, each trail is defined as a quadruple $(\vec{x}_0, \vec{v}_0, \vec{x}_t, \Delta t)$. For this experiment, we use a constant time horizon of $\Delta t = 1s$ for all trails. After visual inspection of the data, the minimal required hit ratio is set to $hit\_ratio_{min}=99.975\%$. We evaluate the models on a random sample of $5 \cdot 10^5$ trails across all three games and all participating players.

\subsection{Preparation of motion models} \label{evaluated_models}

\begin{figure}
    \centering
    \includegraphics[width=\textwidth]{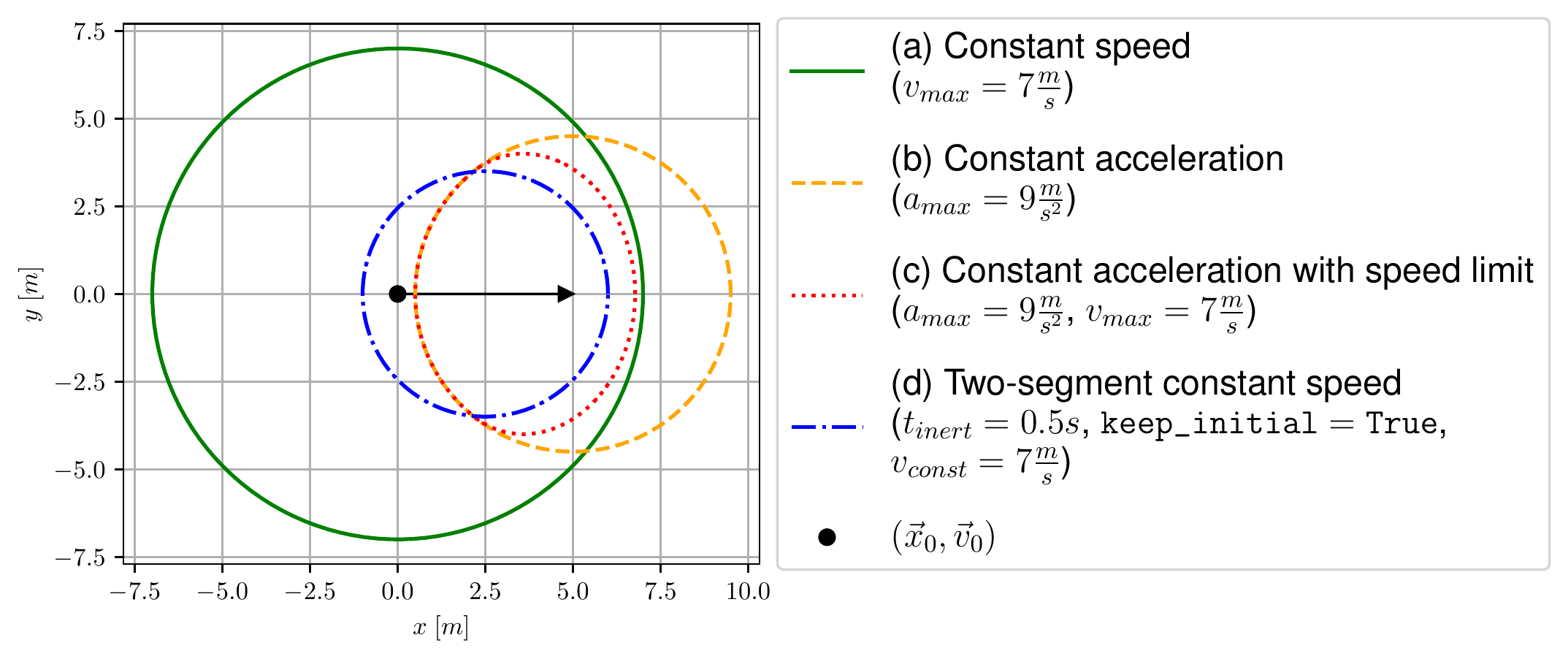}
    \caption{Exemplary boundaries $B$ of the reachable area defined by different motion models when the player starts at $\vec{x}_0 = \vec{0}$ with velocity $\vec{v}_0=\colvec{5\frac{m}{s}}{0}$. The time horizon is $\Delta t = 1s$.}
    \label{fig:reachable_area_per_model}
\end{figure}

We phrase the motion models such that they define a set of reachable positions $R$. This set has to form a non-zero, finite area. To approximate the reachable area as a polygon, $R$ also has to be simply connected and have a computationally approximable boundary $B$. The sets $R$ and $B$ depend on the player's initial position $\vec{x}_0$ and velocity $\vec{v}_0$, the selected time horizon $\Delta t$, and the parameters of the respective model. For brevity of notation, we set $t_0 = 0$ such that $\Delta t = t$. The reachable areas defined by the models (a) - (d) are exemplarily visualized in Figure~\ref{fig:reachable_area_per_model}.

\subsubsection{(a) Constant speed.}

Given motion with some constant speed $v \in [0, v_{max}]$ in direction $\phi$ from a starting location $\vec{x}_0$, the set of reachable target locations $R$ after time $t$ forms a disk with the center $\vec{x}_0$ and radius $v_{max} t$.
\[ R = \{\vec{x} \mid \exists \phi \in [0, 2\pi], \exists v \in [0, v_{max}], \vec{x} = \vec{x}_0 + v \colvec{\cos{\phi}}{\sin{\phi}} t\} \]
\begin{align}
B = \{\vec{x} \mid \exists \phi \in [0, 2\pi], \vec{x} = \vec{x}_0 + v_{max} \colvec{\cos{\phi}}{\sin{\phi}} t\}
\label{eq:B_a}
\end{align}
\subsubsection{(b) Constant acceleration.}

Given motion with some constant acceleration $a \in [0, a_{max}]$ in direction $\phi$ from a starting location $\vec{x}_0$ with starting velocity $\vec{v}_0$, the set of reachable positions $R$ after time $t$ forms a disk with the center $\vec{x}_0 + \vec{v}_0t$ and radius $\frac{1}{2}a_{max}t^2$.
\[ R = \{\vec{x} \mid \exists \phi \in [0, 2\pi], \exists a \in [0, a_{max}], \vec{x} = \vec{x}_0 + \frac{1}{2}a \colvec{\cos{\phi}}{\sin{\phi}} t^2 + \vec{v}_0 t\} \]
\begin{align}
B = \{\vec{x} \mid \exists \phi \in [0, 2\pi], \vec{x} = \vec{x}_0 + \frac{1}{2}a_{max} \colvec{\cos{\phi}}{\sin{\phi}} t^2 + \vec{v}_0 t\}
\label{eq:B_b1}
\end{align}
\subsubsection{(c) Constant acceleration with speed limit.}

We want to restrict model (b) such that the player's speed never exceeds a fixed maximum $v_{max}$. For that purpose, the simulated trajectory is divided into two segments: A first segment of motion with constant acceleration $a$ until $v_{max}$ is reached, and a second segment of motion with constant speed $v_{max}$. A non-negative solution for the time $t_{acc}$ until $v_{max}$ is reached is guaranteed to exist if the player's initial speed $|\vec{v}_0|$ does not exceed $v_{max}$. For that reason, $\vec{v}_0$ is clipped to $\vec{v}_0^* = \vec{v}_0 \frac{\min(|\vec{v}_0|, v_{max})}{|\vec{v}_0|}$. With $t_{acc}^* = \min(t_{acc}, t)$, the reachable area $R$ corresponds to the following simply connected shape, which can be non-circular as $t_{acc}^*$ depends on $\phi$.
\[ R = \{\vec{x} \mid \exists \phi \in [0, 2\pi], \exists a \in [0, a_{max}], \vec{x} = \vec{x}_0 + \vec{v}_0^* t + a t_{acc}^* (t-\frac{1}{2}t_{acc}^*) \colvec{\cos{\phi}}{\sin{\phi}}\} \]
\begin{align}
B = \{\vec{x} \mid \exists \phi \in [0, 2\pi],\vec{x} = \vec{x}_0 + \vec{v}_0^* t + a_{max} t_{acc}^* (t-\frac{1}{2}t_{acc}^*) \colvec{\cos{\phi}}{\sin{\phi}}\} 
\label{eq:B_b2}
\end{align}

\subsubsection{(d) Two-segment constant speed.}

We propose the following approximate model to respect the current kinematic state of a player: The player's motion is divided into two segments of constant-speed motion. First, with some speed $v_{inert}$ for a predetermined amount of time $t_{inert}$ in the direction of $\vec{v}_0$ and, subsequently, with speed $v_{final}$ in some arbitrary direction $\phi$. Using $t_{inert}^* = \min(t_{inert}, t)$, the set of reachable positions $R$ forms a disk with the radius $v_{final}(t-t_{inert}^*)$ if $t_{inert} > t$ and is otherwise reduced to the point $\vec{x}_0 + v_{inert} \frac{\vec{v}_0}{|\vec{v}_0|} t$. 
\[ R = \{\vec{x} \mid \exists \phi \in [0, 2\pi], \exists v \in [0, v_{final}], \vec{x} = \vec{x}_0 + v_{inert} \frac{\vec{v}_0}{|\vec{v}_0|} t_{inert}^* + v_{final} \colvec{\cos{\phi}}{\sin{\phi}}(t-t_{inert}^*) \} \]
\begin{align}
B = \{\vec{x} \mid \exists \phi \in [0, 2\pi], \vec{x} = \vec{x}_0 + v_{inert} \frac{\vec{v}_0}{|\vec{v}_0|} t_{inert}^* + v_{final} \colvec{\cos{\phi}}{\sin{\phi}}(t-t_{inert}^*)
\label{eq:B_c}
\end{align}
The values of $v_{inert}$ and $v_{final}$ can in principle be set in various ways. In our parameterisation, we determine the value of $v_{inert}$ according to a boolean parameter \texttt{keep\_initial}, such that the speed of the player is either preserved or set to a fixed value $v_{const}$.
\[v_{inert} = \begin{cases}
|\vec{v}_0| & \text{if } \texttt{keep\_initial} \\
v_{const} & \, \text{else}
\end{cases}\]
The value of $v_{final}$ is either computed based on two parameters $a_{max}$ and $v_{max}$ which corresponds to pretending that the player has accelerated during the first segment, or set to the fixed value $v_{const}$.
\[v_{final} = \begin{cases}
\min(v_{inert} + a_{max}t_{inert}, v_{max}) & \text{if } a_{max} \text{ and } v_{max} \text{ are set}\\
v_{const} & \, \text{else}
\end{cases}\]

\subsubsection{Computation of reachable area}

It follows from equations~\eqref{eq:B_a} - \eqref{eq:B_c} that for each presented model, the reachable area can be approximated as a polygon by computing the vertices of that polygon as points from the boundary $B$ along a discrete set of angles $\phi$. For our evaluation, we choose to evaluate $200$ evenly spaced angles.

\subsection{Optimisation of model parameters} \label{optimisation}
Based on the validation procedure outlined in \cref{validation_procedure}, we aim to find the motion model $m$ out of the set of considered models $M$ which maximises the $score$ \eqref{eq_validation_score} for a given sample of trails $T$. We define the optimal model as:
\begin{equation} \label{arg_max_compact}
    \underset{m \in M}{\operatorname{arg \: max}}  \ (score(m, T))
\end{equation}
However, for each of the four player motion models defined in \cref{evaluated_models}, there are free parameters which specify its behaviours. For this reason, in order to determine the best model, the optimal combination of parameters has to be found for each model. We thus extend \eqref{arg_max_compact} by introducing the tuple $P_m$ that represents the values for the free parameters of a model $m(P_m)$:
\begin{equation}
    \underset{m \in M}{\operatorname{arg \: max}}\ (
        \underset{P_m}{\operatorname{max}}  \ (score(m(P_m), T))
    )
\end{equation}
The parameters of the models (a) - (d) are, as outlined in \cref{evaluated_models}:

\begin{itemize}
    \item (a) Constant speed: $P_{(a)} = (v_{max})$
    \item (b) Constant acceleration: $P_{(b)} = (a_{max})$
    \item (c) Constant acceleration with speed limit: $P_{(c)} = (a_{max}, v_{max})$
    \item (d) Two-segment constant speed: $P_{(d)} = (t_{inert}, \texttt{keep\_initial}, v_{const}, a_{max}, v_{max})$
\end{itemize}

The main challenge here is to determine the optimal $P_m$ for a model with regard to its $score$ \eqref{eq_validation_score}. For this experiment, the best parameter configuration $P_m$ is determined using Bayesian optimisation. However, Bayesian optimization generally does not work for discrete parameter values such as the boolean variable $\texttt{keep\_initial}$ in model (d). There are proposals for enabling discrete variables, most notably by Luong et al. \cite{luong_bayesian_2019}, but this works only if all values in $P_m$ are discrete, which does not apply to our models.
Luckily, our models have very few discrete parameters, so we perform one Bayesian optimisation for each combination of discrete parameter values and use the best score across those results.

\subsection{Evaluation of results}

\begin{figure}
    \centering
    \includegraphics[width=\textwidth]{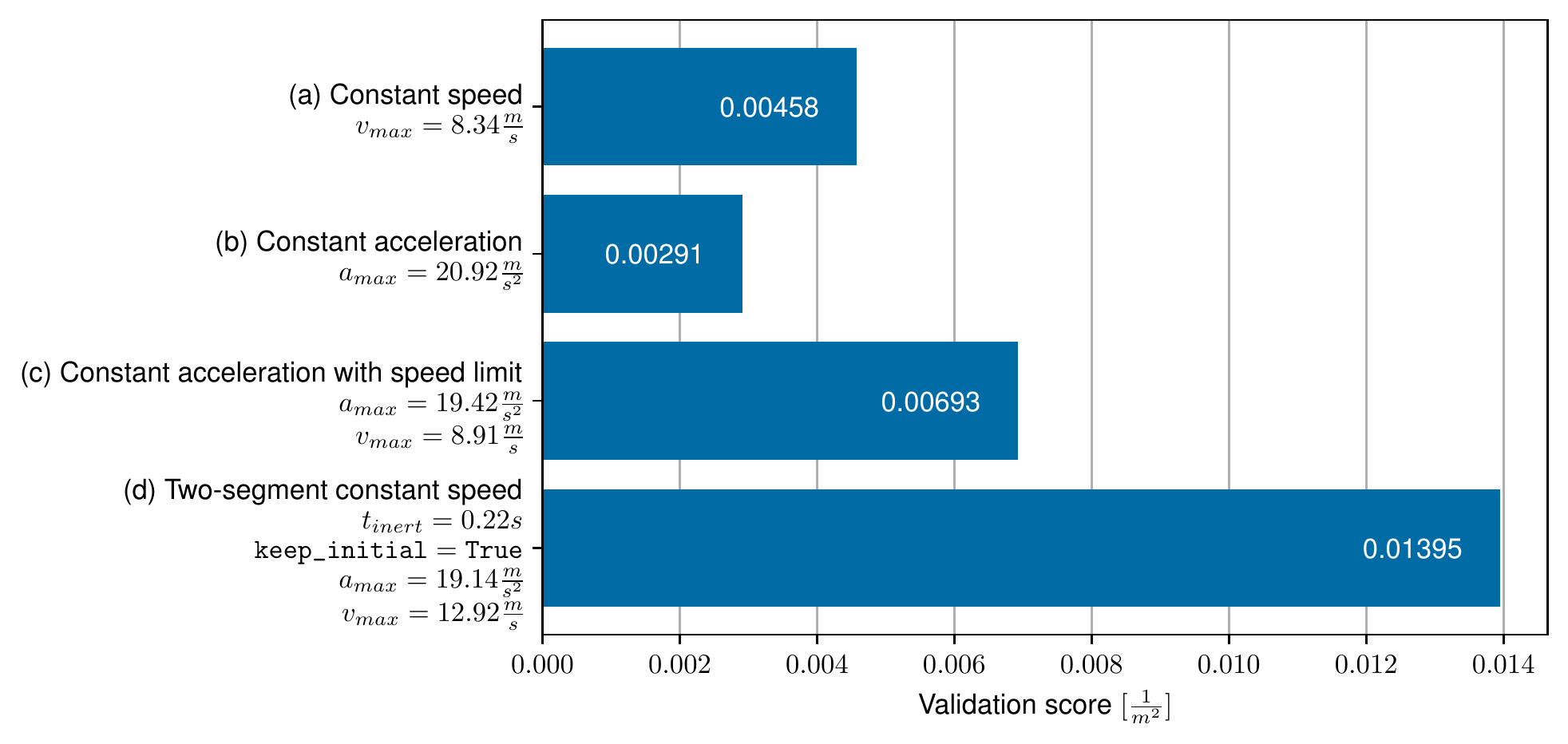}
    \caption{Comparison of the performance of the four models (a) - (d) with their optimised parameter values.}
    \label{fig:results}
\end{figure}

The performance of the optimised models (a) - (d) and their parameter values are shown in Figure~\ref{fig:results}.

The constant-speed model (a) unsurprisingly shows a weaker performance ($score^{-1}=218m^2$) than the more sophisticated models (c) and (d), since it does not factor in the initial kinematic state of the player. If the player moves at a high speed, the model unrealistically assumes that the player can instantly run at maximal speed in the other direction. Thus, for high speeds, the model overestimates the amount of reachable space in directions which sufficiently deviate from the direction which the player is initially moving towards.

The naive constant acceleration model (b) ($score^{-1}=344m^2$) performs even worse than model (a), likely because it makes the unrealistic assumption that the possible magnitude of acceleration is independent of the magnitude and direction of a player's current velocity. This implies in particular that for high speeds, the amount of reachable space in the direction that a player is moving towards will be heavily overestimated since the model assumes that the player's speed can increase unboundedly.

The model assuming constant acceleration with a speed limit (c) ($score^{-1}=144m^2$) outperforms models (a) and (b). However, the optimised value of the maximally possible acceleration of a player $a_{max}$ is physically unrealistic. A value of $a_{max}=19.42 \frac{m}{s^2}$ assumes that a player can accelerate from zero to the top speed $v_{max}=8.91\frac{m}{s} (=32.08\frac{km}{h})$ within about half a second, which is implausibly fast. Therefore, the model still overestimates the reachable area. One approach to improve the constant acceleration model could be to view the maximally possible acceleration as being dependent on the direction and magnitude of the current velocity.

The two-segment constant speed model (d) ($score^{-1}=71.7m^2$) is able to account for all reachable positions by predicting only about half the area of model (c). It successfully narrows down the area that a player can reach within one second to a circle with an average radius of $4.8$ meters which is highly accurate. Model (d) not only achieves the best score in our evaluation, but is also mathematically simpler than model (c). For that reason, it is also computationally more efficient across the various tasks that motion models are used for, like the computation of reachable areas or the shortest time to arrive at a specific location. A drawback of model (d) is that in its presented form, it only makes meaningful predictions for time horizons $\Delta t > t_{inert}$ (where the optimised value is $t_{inert} = 0.22s$). Below this duration, the model is not applicable unless it is appropriately extended.

In summary, our newly presented model (d) is both highly accurate and computationally efficient. The models (a) - (c) have obvious weaknesses and are not adequate to accurately identify reachable locations.

\section{Conclusion}\label{conclusion}

We presented a novel approach to the validation and optimisation of models of trajectorial player motion in football and similar sports. We also presented an empirical comparison of the accuracy of various such models. In line with our expectations, more sophisticated and accurate assumptions made by a motion model generally tend to be reflected in a better predictive performance. Yet, by far the best-performing model is our proposed approximate model which assumes motion along two segments with constant speed. Using this model allows researchers to compute complex performance indicators more efficiently and accurately over large data sets. In contrast, player motion should not be assumed to take place with constant speed or constant acceleration with unlimited speed. These assumptions are inappropriate to accurately distinguish reachable from unreachable locations.

The validation and optimisation approach described in this paper can be applied to data with arbitrary distributions of measurement error. However, this is also a disadvantage, since the threshold for the amount of outliers that are attributed to measurement error has to be determined manually. This threshold also has to be set for each distinguished population, depending on the frequency of extrema and the distribution of measurement error in the population. If for example one wants to evaluate motion models for each player individually, it would be misleading to assign the same threshold to each player because players who sprint regularly during the game produce more positional extrema (and thus outliers) than goalkeeperes, for example, who are rarely forced to run with maximal effort. A solution to the problem of having to specify a threshold is to perform validation and optimisation with different thresholds and analyse how this choice affects the result. Also, the approach presented here can be extended to automatically determine an optimal threshold for known error distributions.

In the future, we plan to search for motion models that further exceed the presented ones in accuracy and computational efficiency. A key towards this goal is to estimate motion models from positional data. Many problems addressed in this paper are mirrored in empirical model fitting, for example the need to exclude outliers and the lack of generalisability across populations~\cite{brefeld2019probabilistic}. In the context of validation, empirical models can serve as a highly informative benchmark to reveal how well theoretical models are able to approximate actual human motion.

\bibliographystyle{splncs04}
\bibliography{references}

\clearpage

\end{document}